\documentclass[12pt,a4paper]{article}
 \usepackage{epsf}
\begin{document}
 \title{Neutral top-pion and top-charm production in high energy
 $e^{+}e^{-}$ collisions}
 \author{Chongxing Yue$^{(a,b)}$, Gongru Lu$^{(a,b)}$, Junjie Cao$^{b}$,
 Jiantao Li$^{b}$, Guoli Liu$^{b}$ \\
 {\small b.College of Physics and Information Engineering},\\
 {\small Henan Normal University,
 Xinxiang 453002. P.R.China}
 \thanks{This work is supported by the National Natural Science
 Foundation of China, the Excellent Youth Foundation of Henan Scientific
 Committee; and Foundation of Henan Educational Committee.}
 \thanks{E-mail: cxyue@pbulic.xxptt.ha.cn} }
 \date{\today}
 \maketitle
 \begin{abstract}
 \hspace{5mm} We calculate the contributions of the
 neutral top-pion, predicted by topcolor-assisted technicolor
 (TC2) theory, to top-charm production via the processes
 $\gamma\gamma\longrightarrow\overline{t}c$ and
 $e^{+}e^{-}\longrightarrow \gamma\gamma\longrightarrow
 \overline{t}c$ at the high energy linear
 $e^{+}e^{-}$ collider (LC) experiments. The cross section is of
 order $10^{-2}pb$ in most of the parameter space of TC2 theory,
 which may be detected at the LC experiments.
 So the process $e^{+}e^{-}\longrightarrow \overline{t}c$
 can be used to detect the signature of TC2 theory.
 \end {abstract}
 \vspace{1.0cm}

 \newpage
 It is often stated that the large value of the top-quark mass
 opens the possibility that it plays a special role in current
 particle physics. Indeed, the properties of top quark could reveal
 information on flavor physics, electroweak symmetry breaking (EWSB)
 as well as new physics beyond standard model. One of these
 consequences is that the flavor changing scalar coupling (FCSCs) at
 the tree level would exist at high mass scales. The measurement of
 FCSCs involving a top quark would provide an important test for
 the various beyond standard models \cite{y1,y2}.

 Topcolor-assisted technicolor (TC2) theory \cite{y3} is an
 attractive scheme in which there is an explicit dynamical
 mechanism for breaking electroweak symmetry and generating
 the fermion masses including the heavy top quark mass
 $m_{t}=175GeV$. In TC2 theory, the EWSB is driven mainly by
 technicolor (TC)interactions, the extended technicolor (ETC)
 interactions give contributions to all ordinary quark and lepton
 masses including a very small portion of the top quark mass,
 namely $m_t^{\prime}=\varepsilon m_{t}$ \cite{y4} with a model-dependent
 parameter $\varepsilon$ $(0.03{\leq}\varepsilon{\leq} 0.1)$. The
 topcolor interactions also make small contributions to the EWSB
 and give rise to the main part of the top quark mass
 $m_{t}-m_{t}^{\prime}=(1-\varepsilon)m_{t}$ similar to the constituent
 masses of the light quarks in QCD. This means that the associated
 top-pions $\pi_{t}^{0}$, $\pi_{t}^{\pm}$ are not the longitudinal
 bosons $W$ and $Z$, but separately, physically observable objects.
 Thus top-pions can be seen as the characteristic feature of TC2
 theory. Studying the possible signatures of top-pions at high
 energy colliders can be used to test TC2 theory.

 Based upon the observation that the couplings of the charged
 top-pions $ \pi_{t}^{\pm}$ to charm and bottom quarks can be large
 due to a significant mixing of the top and charm quarks, H.-J.
 He and C.-P Yuan \cite{y5} have studied the possibility of
 detecting the charged top-pions $\pi_{t}^{\pm}$ at colliders. In
 Ref.[6], G.Burdman has considered the prospects of the
 observation of the neutral top-pion $\pi_{t}^{0}$ via the process
 $gg{\longrightarrow}\overline{t}c$ at hadron colliders. In this letter,
 we consider the process $\gamma\gamma{\longrightarrow}
 \pi_{t}^{0}\longrightarrow\overline{t}c$ and see
 whether $\pi_{t}^{0}$ can be detected via this process at high
 energy linear $e^{+}e^{-}$ collider (LC) experiments. Our results
 show that this process is important in probing the flavor-charging
 top-charm-scalar couplings. If the parameter values of TC2 theory
 are favorable, the neutral top-pion $\pi_{t}^{0}$ may be
 detectable at the LC experiments.

 The possibility of transforming a high energy $e^{+}e^{-}$ collider
 into a photon-photon $(\gamma\gamma)$ collider has deserved a lot
 of attention. For example, Ref.[7] has showed that the top quark
 pair production in $\gamma\gamma$ colliders is larger than the
 direct $e^{+}e^{-}\longrightarrow t\overline{t}$ production both
 with and without considering the threshold QCD effect. And Ref.[8]
 has pointed out that the difference beyond standard models, such
 as the minimal supersymmetric standard model and various TC
 models, can be distinguished by the process
 $\gamma\gamma{\longrightarrow}t\overline{t}$ cross section
 measurement at the high-energy $\gamma\gamma$ colliders. Ref.[2]
 has showed that the two-Higgs-doublet model which induces FCSCs is
 more naturally probed via the process $\gamma\gamma\longrightarrow
 \overline{t}c$ than in flavor changing top-quark
 decays due to the large underlying mass scales and possibly large
 momentum transfer. The neutral top-pion, as an isospin triplet, can
 couple to the photons via the top quark triangle loop in an
 isospin violating way similar to the coupling of QCD pion
 $\pi^{0}$ to the gluons, and the large isopion violation $(m_{t}-
 m_{b})/(m_{t}+m_{b})\approx 1$ makes its contribution to the top
 quark production cross section very important \cite{y9}. Thus, in
 this letter we consider the process
 $\gamma\gamma\longrightarrow \pi^{0}\longrightarrow \overline{t}c$
 and see whether the $\pi_{t}^{o}$ exchange can
 mediate top-charm production at interesting levels at the LC
 experiments.

 For TC2 theory, the TC interactions play the major role in
 breaking the electroweak gauge symmetry, while the topcolor
 interactions also make small contributions to the EWSB. Thus,
 there is the following relation:
 \begin{equation}
 (\nu_{\pi})^{2}+(F_{t})^{2}=(\nu_{w})^{2}
 \end{equation}
 where $\nu_{\pi}$ represents the contributions of TC interactions
 to the EWSB, $F_{t}=50GeV$ is the top-pion decay constant, and
 $\nu_{w}= \nu/\sqrt{2}=174GeV$.

 For TC2 theory, it generally predicts three light top-pions with
 large Yukawa coupling to the third family. This induces distinct
 new flavor changing scalar couping. The relevant flavor changing
 scalar couplings including the $t-c$ transition for the neutral
 top-pion can be written as \cite{y3,y5,y6}:
 \begin{equation}
 {m_{t}\over\sqrt{2}F_{t}}
 {\sqrt{\nu_{w}^{2}-F_{t}^{2}}\over\nu_{w}}
 [k_{UR}^{tt}k^{tt^{*}}_{UL}\overline{t}_{L}t_{R}\pi_{t}^{0}+k_{UR}^{tc}
 k^{tt^{*}}_{UL}\overline{t}_{L} c_{R} \pi_{t}^{0}+h.c.]
 \end{equation}
 where the factor ${\sqrt{\nu_{w}^{2}-F_{t}^{2}}/\nu_{w}}$
 reflects the effect of the mixing between the neutral top-pion
 $\pi_{t}^{0}$ and the would be Goldston boson \cite{y10}.
 $k^{ij}_{UL}$ is the matrix element of unitary matrix $k_{UL}$
 which the CKM matrix can be derived as $V=k_{UL}^{-1} k_{DL}$ and
 $k_{UR}^{ij}$ is the matrix element of the right-handed relation
 matrix $K_{UR}$. Ref.[5] has shown that their values can be taken
 as:
 \begin{equation}
 k_{UL}^{tt}=1,\hspace{5mm} k^{tt}_{UR}=1-\epsilon,\hspace{5mm}
 k_{UR}^{tc}\leq \sqrt{2\varepsilon-\varepsilon^{2}}
 \end{equation}
 In the following calculation, we will take
 $k^{tc}_{UR}=\sqrt{2\varepsilon-\varepsilon^{2}}$.

 For the top quark triangle loop, the simple Adler-Bell-Jackiw (ABJ)
 anomaly approach is not sufficient since the top quark mass is
 only $175GeV$. Here, we explicitly calculate the top quark
 triangle loop and obtain the following $\pi_{t}^{0}-\gamma-\gamma$
 coupling:
 \begin{equation}
 \frac{- N_{C}e^{2}(m_{t}-m_{t}^{'})m_{t}}{12\sqrt{2}\pi^{2}F_{t}}
 C_{0} \pi _{t}^{0}
 \epsilon_{\mu\nu\lambda\rho}(\partial^{\mu}A^{\nu})
 (\partial^{\lambda}A^{\rho})
 \end{equation}
 where $N_{C}$ is the color index with $N_{C}=3,\hspace{3mm}
 C_{0}(p_{4},-p_{4},-p_{3},m_{t},m_{t},m_{t},)$ is the standard
 three-point Feynman integral with $p_{3}$ and $p_{4}$ donating the
 momenta of the two incoming photons.

 Ref.[3, 4] have estimated the mass of the top-pion in the fermion
 loop approximation and given $180GeV\leq m_{\pi_t} \leq 250GeV$
 for $m_t=180GeV$ and $0.03\leq \varepsilon \leq 0.1$. Since
 the negative top-pion corrections to the $z \rightarrow b\bar{b}$
 branching ratio $R_b$ become smaller when the top-pion is
 heavier, the $LEP/SLD$ data of $R_b$ give rise to certain lower
 bound on the top-pion mass. It was shown in Ref.[10] that the
 top-pion mass should not be lighter than the order of $1TeV$
 to make the TC2 theory consistent with the $LEP/SLD$ data.
 However, we restudied this problem in Ref.[11]. Our results
 show that the top-pion mass is allowed to be in the region of a
 few hundred $GeV$ depending on the models. Thus the top-pion mass
 depends on the value of the parameters in the TC2 models. As
 estimation the contributions of the neutral top-pion $\pi_t^0$
 to $\bar{t}c$ production, we take the mass of the $\pi_t^0$
 to vary in the range of $200GeV-350GeV$ in this letter.

 Using the above formula, we can obtain the cross section
 $\widehat{\sigma}(\widehat{s})$ of the process
 $\gamma\gamma\longrightarrow  \pi^{0}_{t}\longrightarrow
 \overline{t}c$ which can be written as:
 \begin{equation}
 \widehat{\sigma}(\widehat{s})=\frac{3
 e^{4}A^{2}}{32\pi}\frac{\widehat{s}
 (\widehat{s}-m_{t}^{2})^{2}}{\widehat{s}+m_{t}^{2}}
 \end{equation}
 with
 \begin{equation}
 A=\frac{m_{t}^{3}(1-\epsilon) k^{tc}_{UR}}{6\pi ^{2}F_{t}^{2}}
 \sqrt{\frac{(\nu_{w}^{2}-F_{t}^{2})}{\nu_{w}^{2}}}
 \frac{1}{\widehat{s}-m_{\pi_{t}}^{2}+im_{\pi_t}\Gamma_{total}}C_{0}
 \end{equation}
 where $\sqrt{\widehat{s}}$ is the $\gamma\gamma$ center-of-mass
 energy. $\Gamma_{total}$ is the total decay width of $\pi^0_t$.
 The possible decay modes of $\pi_t^0$ are $b\bar{b}$,
 $\bar{t}c$, $gg$, $\gamma\gamma$, $zz$ and $z\gamma$ for
 $200GeV\leq m_{\pi_t} \leq 350GeV$. Our calculation results show
 that the branching ratio $Br(\pi_t^0 \rightarrow \bar{t}c)$ is
 larger than $85\%$ for $\varepsilon\geq 0.01$ and is larger than
 $98\%$ for $\varepsilon\geq 0.03$. In this letter, we assume that
 the free parameter $\varepsilon$ is in the region of $0.03\sim
 0.1$, so the total width $\Gamma_{total}$ can be written as :
 \begin{eqnarray}
 \Gamma_{total} &\simeq & \Gamma(\pi_t^0 \rightarrow \bar{t}c)
 \\ \nonumber
 &\simeq & \frac{3(1-\varepsilon)^2}{16\pi}
 \frac{\nu_w^2-F_t^2}{\nu_w^2}
 \frac{m_t^2 m_{\pi_t}}{F_t^2}(K_{UR}^{tc})^2
 \sqrt{1-\frac{m_t^2}{m_{\pi_t}^2}}
 \end{eqnarray}

 The total cross section $\sigma(s)$ of the process $e^{+}e^{-}
 \longrightarrow \gamma\gamma \longrightarrow \overline{t}c$ can be
 obtained by folding the cross section
 $\widehat{\sigma}(\gamma\gamma\longrightarrow \overline{t}c)$
 with the photon luminosity at the $e^{+}e^{-}$ colliders
 \cite{y12,y8}:
 \begin{equation}
 \sigma(s)=\int_{(m_{t}+m_{c})/\sqrt{s}}^{x_{max}}(dz)
 (dL_{\gamma\gamma})/(dz)
 \widehat{\sigma}(\widehat{s}) \hspace{8mm} (\widehat{s}=z^{2}s)
 \end{equation}
 where $\sqrt{s}$ is the $e^{+}e^{-}$ centre-of-mass energy and
 $(dL)_{\gamma\gamma}/(dz)$ is the photon luminosity which is given in
 Ref.[12].

 According to the method used in Refs.[12, 8], we can calculate
 the subprocess cross section $\widehat{\sigma}(\widehat{s})$ and
 cross section $\sigma(s)$. Fig.1 shows the cross sections
 $\widehat{\sigma}(\widehat{s})$ for the process
 $\gamma\gamma\longrightarrow \overline{t}c$ as a
 function of the mass of neutral top-pion $\pi^{0}_{t}$ at
 $\sqrt{\widehat{s}} =300GeV$. The cross sections are displayed for
 the three values of the parameter $\varepsilon=0.05, 0.08,$ and
 $0.1$, respectively. From Fig.1, we can see that the cross sections
 are not sensitive to the value of parameter $\varepsilon$. The peak
 of each cure emerges at $m_{\pi_{t}}\sim \sqrt{\widehat{s}}$. The
 maximum value can reach $1pb$ for $m_{\pi_{t}}=300GeV$,
 $\varepsilon=0.05$.

 To see the effect of the center-of-mass energy $\sqrt{\widehat{s}}$
 on the top-charm production rate, we plot the cross section
 $\widehat{\sigma}( \widehat{s})$ as functions of
 $\sqrt{\widehat{s}}$ for $\varepsilon=0.08$ in Fig.2, in which the
 solid, dotted and dashed lines stand for $m_{\pi_{t}}=200GeV,
 250GeV$ and $300GeV$, respectively. From these curves we find that
 the cross section can be obviously enhanced when $m_{\pi_{t}}$ get
 close to $\sqrt{\widehat{s}}$. $\widehat{\sigma}( \widehat{s})$ is
 larger than $10^{-2}pb$ for $m_{\pi_t}\geq 250GeV$.

 In Fig.3, we show the cross section $\sigma(s)$ of the process
 $e^{+}e^{-}\longrightarrow \gamma\gamma\longrightarrow
 \overline{t}c$ as a function of center-of-mass
 energy of electron-positron system $\sqrt{s}$ with
 $\varepsilon=0.08$, $m_{\pi_{t}}=200GeV, 250GeV$ and $300GeV$.
 From Fig.3 we can see that the cross section $\sigma(s)$ of the
 process $e^{+}e^{-}\longrightarrow \gamma\gamma\longrightarrow
 \overline{t}c$ is very small for $\sqrt{s}<230GeV$ and the lines
 go to flat as $\sqrt{s}$ increasing. This is due to : (a) the
 available range of $\sqrt{\widehat{s}}$ is determined by
 $min(\sqrt{\widehat{s}})=m_t+m_c\simeq 177GeV$ and
 $max(\sqrt{\widehat{s}})=x_{max}\sqrt{s}=0.83\sqrt{s}$,
 so that $max(\sqrt{\widehat{s}})$ is $s$ dependent; (b)
 the $\gamma\gamma$ luminosity $(dL)_{\gamma\gamma}/(dz)$
 decreases rapidly in the vicinity of $max(\sqrt{\widehat{s}})$
 which give rise to a suppression of the large
 $\sqrt{\widehat{s}}$ contributions to $\sigma(s)$;
 (c) the integrated range of the $\sigma(s)$ is in the range of
 $(m_t+m_c)/\sqrt{s}\sim x_{max}$ which increase as $\sqrt{s}$
 increasing.

 To see the effect of neutral top-pion mass on the cross section
 $\sigma(s)$, we plot $\sigma(s)$ versus as function of
 $m_{\pi_{t}}$ in Fig.4 for $\varepsilon=0.05, 0.08, 0.1$ and
 $\sqrt{s}=500GeV$. From Fig.4 we can see that $\sigma(s)$ is not
 sensitive to the parameter $\varepsilon$ and increases with
 $m_{\pi_{t}}$ increasing. The cross section $\sigma(s)$ varies
 between $8.5\times10^{-3}pb$ and $0.146pb$ as $m_{\pi_{t}}$
 increases from $200GeV$ to $350GeV$. The reason is that the
 available $\sqrt{\widehat{s}}$ spreads in a wider range of
 $177GeV\sim 415GeV$ which increase the importance of the
 contributions of the factor $(\widehat{s}-m_{\pi_t}^2+
 im_{\pi_t}\Gamma_{total})^{-1}$ and the $\gamma\gamma$ luminosity
 suppression effect is less significant in this range.

 The cross section of the process $e^{+}e^{-}\longrightarrow
 \overline{t}c$ is very small at one loop level in
 the standard model due to the unitary of CKM matrix. New particles
 predicted by the beyond standard may have significant
 contributions to the process $e^{+}e^{-}\longrightarrow
 \overline{t}c$ . Therefore this process can be used
 to test new physics. In this letter, we have discussed and
 calculated the contribution of the neutral top-pion $\pi ^{0}_{t}$
 to the top-charm production via the processes
 $\gamma\gamma\longrightarrow \overline{t}c$ and
 $e^{+}e^{-}\longrightarrow \gamma\gamma \longrightarrow
 \overline{t}c$ at the LC experiments. We find that
 the cross section is of order $10^{-2}pb$ in the reasonable
 parameter space of TC2 theory. If we assume that the integrated
 luminosity of high energy phone-phone ($\gamma\gamma$) or
 electron-positron ($e^{-}e^{+}$) colliders is $50fb^{-1}$, it
 would produce a few hundred events at the LC experiments. So it is
 possible to detect the signature of TC2 theory via the processes
 $e^{+}e^{-}\longrightarrow\overline{t}c$ or
 $\gamma\gamma \longrightarrow\overline{t}c$ at the
 future LC experiments.
 \newpage
 \vskip 2.0cm
 \begin{center}
 {\bf Figure Captions}
 \end{center}
 \begin{description}
 \item[Fig.1:]The cross section $\widehat{\sigma}(\widehat{s})$ of the process
 $\gamma\gamma\longrightarrow \pi_{t}^{0}\longrightarrow
 \overline{t}c$ versus $m_{\pi_{t}}$  for
 $\sqrt{\widehat{s}} =300GeV$ and $\varepsilon=0.05,0.08,0.1$.
 \item[Fig.2:]The cross section $\widehat{\sigma}(\widehat{s})$ versus
 $\sqrt{\widehat{s}}$ for $\varepsilon=0.08$ and $m_{\pi_{t}} =200GeV,
 250GeV, 300GeV$.
 \item[Fig.3:]The cross section $\sigma(s)$ of the process $e^{+}e^{-}\longrightarrow
 \gamma\gamma\longrightarrow \overline{t}c$ versus
 $\sqrt{s}$ for $\varepsilon=0.08$ and $m_{\pi_{t}}=200GeV, 250GeV,
 300GeV$.
 \item[Fig.4:]The cross section $\sigma(s)$ versus $m_{\pi_{t}}$ for
 $\sqrt{s}=500GeV$ and $\varepsilon=0.05, 0.08, 0.1$.
 \end{description}
 
 
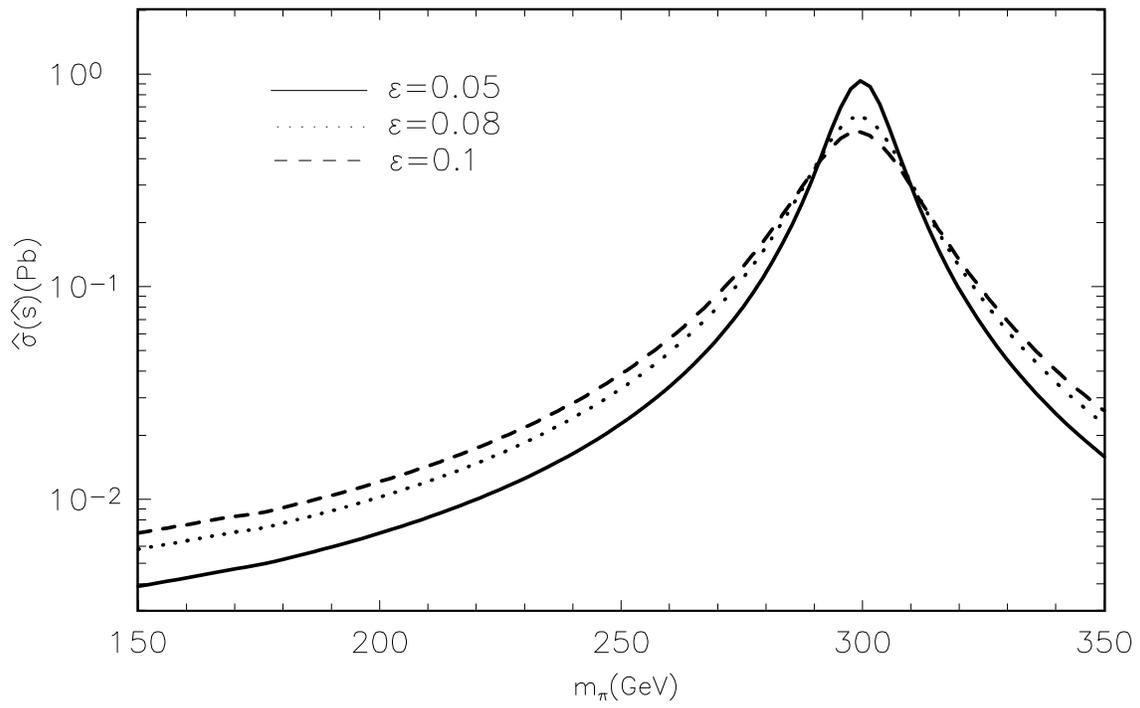
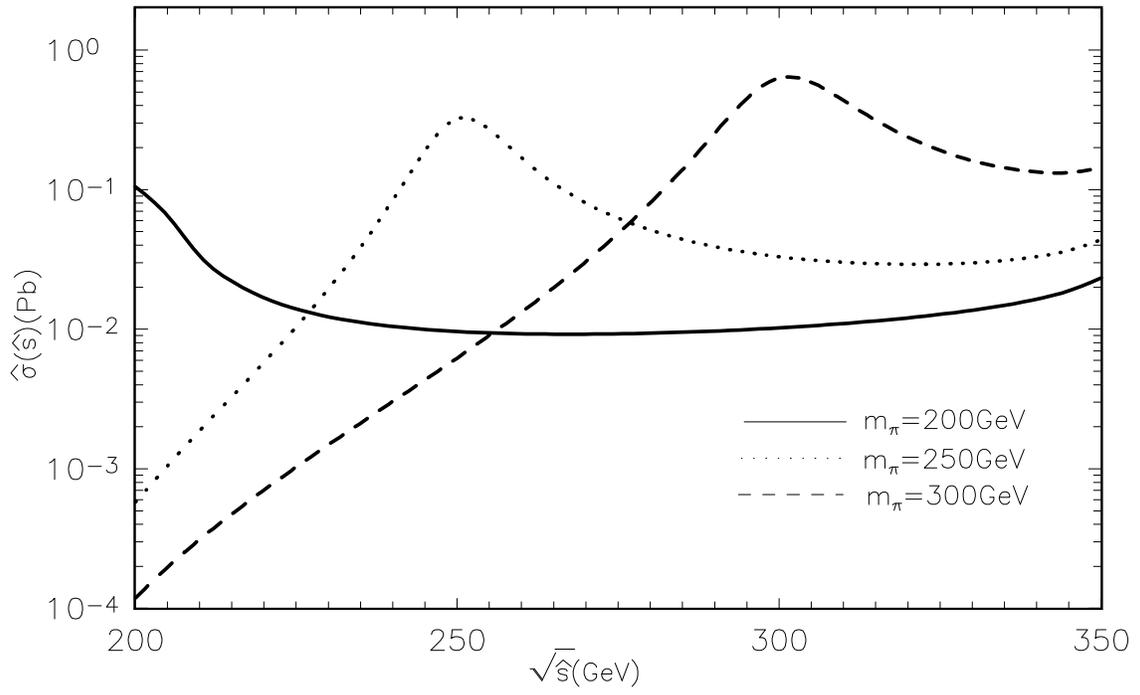
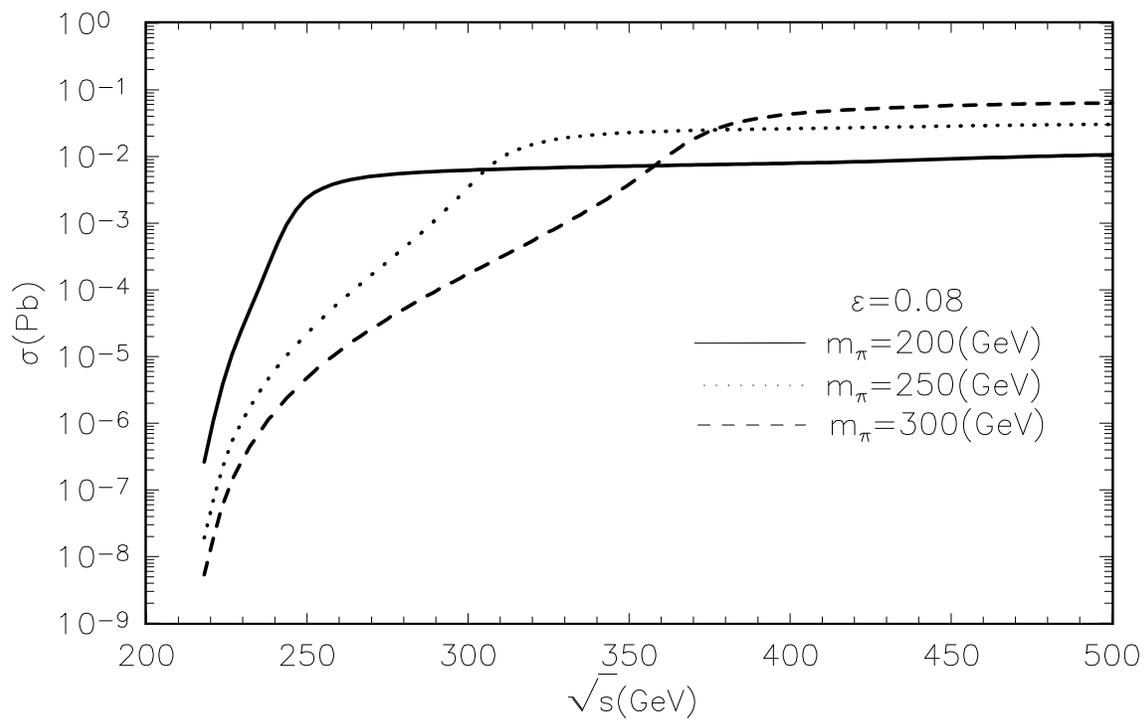
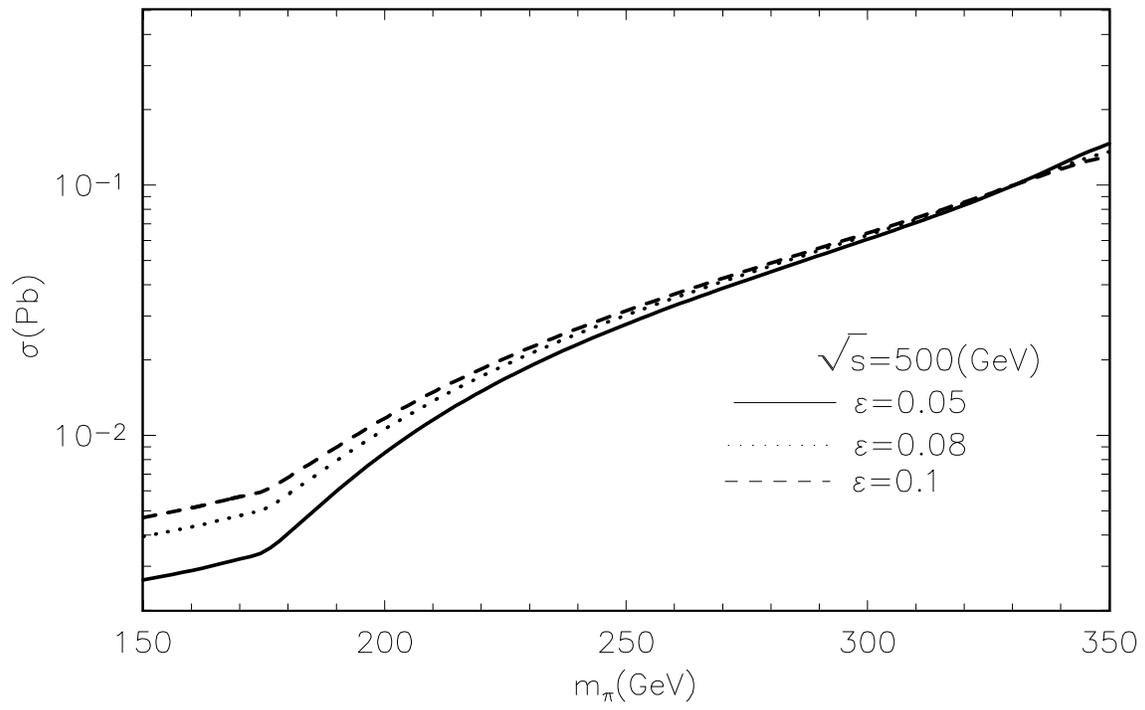
\end{document}